\documentclass[aps,prl,twocolumn,showpacs,superscriptaddress,floatfix]{revtex4-1}
\usepackage{amsmath,amssymb}
\usepackage{graphicx}
\usepackage{epsfig}
\usepackage{color}
\usepackage{rotating}
\usepackage{bm}
\usepackage{setspace}
\begin{document}
\title{Observation of Double-Dome Superconductivity in Potassium-Doped FeSe Thin Films}
\author{Can-Li Song}
\affiliation{State Key Laboratory of Low-Dimensional Quantum Physics, Department of Physics, Tsinghua University, Beijing 100084, China}
\affiliation{Collaborative Innovation Center of Quantum Matter, Beijing 100084, China}
\author{Hui-Min Zhang}
\author{Yong Zhong}
\author{Xiao-Peng Hu}
\affiliation{State Key Laboratory of Low-Dimensional Quantum Physics, Department of Physics, Tsinghua University, Beijing 100084, China}
\author{Shuai-Hua Ji}
\author{Lili Wang}
\author{Ke He}
\author{Xu-Cun Ma}
\email[]{xucunma@mail.tsinghua.edu.cn}
\author{Qi-Kun Xue}
\email[]{qkxue@mail.tsinghua.edu.cn}
\affiliation{State Key Laboratory of Low-Dimensional Quantum Physics, Department of Physics, Tsinghua University, Beijing 100084, China}
\affiliation{Collaborative Innovation Center of Quantum Matter, Beijing 100084, China}

\begin{abstract}
We report on the emergence of two disconnected superconducting domes in alkali-metal potassium (K)-doped FeSe ultra-thin films grown on graphitized SiC(0001). The superconductivity exhibits hypersensitivity to K dosage in the lower-$T_\textrm{c}$ dome, whereas in the heavily electron-doped higher-$T_\textrm{c}$ dome it becomes spatially homogeneous and robust against disorder, supportive of a conventional Cooper-pairing mechanism. Furthermore, the heavily K-doped multilayer FeSe films all reveal a large superconducting gap of $\sim$ 14 meV, irrespective of film thickness, verifying the higher-$T_\textrm{c}$ superconductivity only in the topmost FeSe layer. The unusual finding of a double-dome superconducting phase has stepped towards the mechanistic understanding of superconductivity in FeSe-derived superconductors.

\end{abstract}

\maketitle
%\begin{spacing}{0.9931}
\begin{spacing}{1.01}
Interface-enhanced high temperature superconductivity in single-layer FeSe films on SrTiO$_3$ (FeSe/SrTiO$_3$)\cite{qing2012interface, wen2014direct} was discovered in 2012 and exhibits an unexpected high transition temperature $T_\textrm{c}$ from 65 K \cite{liu2012electronic, he2013phase, tan2013interface, peng2014measurement, zhang2015onset} to even 109 K \cite{ge2014superconductivity}. So far, this subject has experienced a tremendous burst of theoretical and experimental activities in the superconductivity community \cite{wen2014direct, liu2012electronic, he2013phase, tan2013interface, peng2014measurement, zhang2015onset, ge2014superconductivity, liu2012atomic, xiang2012high, bang2013atomic, coh2015large, zhang2014interface, lee2014interfacial, deng2014meissner, cui2015interface, Huang2005revealing}, because it offers an unprecedented opportunity in quest of the mysterious mechanism behind Cooper pairing in high-$T_\textrm{c}$ superconductors \cite{qing2012interface}. Using angle resolved photoemission spectroscopy (ARPES) technique, it has been immediately confirmed that the superconducting single-layer FeSe/SrTiO$_3$ films are sized of a rather simple Fermi surface topology with only electron-like band(s) around the zone corner M \cite{liu2012electronic, he2013phase, tan2013interface, lee2014interfacial}. This has posed a massive challenge to the ever prevailing pairing paradigm of iron-based superconductors (Fe-SCs), in which the repulsive interband interaction between the hole-like bands around the zone center $\Gamma$ and electron bands around M leads to strong spin fluctuation and consequently a sign-reversing $s$-wave state ($s_\pm$ pairing symmetry) \cite{Unconventional2008Mazin, Unconventional2008Kuroki, hirschfeld2011gap}. Subsequent theoretical efforts to tentatively interpret this unwonted phenomenon have extended the $s_\pm$ pairing model and generated ``incipient'' $s_\pm$ wave, nodeless \textit{d} wave and a more subtle sign-changed \textit{s}-wave state between two M-near hybridized electron pockets \cite{hirschfeld2011gap}, but receiving little attention. Alternatively, the conventional Cooper pairing mechanism based on phonon scenario from either FeSe itself \cite{xiang2012high, coh2015large} or cross the interface \cite{lee2014interfacial, cui2015interface, lee2015makes}, in conjunction with electron transfer from SrTiO$_3$ to FeSe films, has been proposed and attracts increasing attention \cite{qing2012interface, lee2014interfacial, lee2015makes}. This is further supported by the recent observation of plain \textit{s}-wave superconductivity with no sign change, rather than spin fluctuation driven $s_\pm$ wave state and its extensions in  FeSe/SrTiO$_3$ \cite{fan2015plain}.

On the other hand, the recent demonstrations of high temperature superconductivity in heavily electron-doped FeSe films/flakes through alkali-metal potassium (K) \cite{miyata2015high, wen2015anomalous, tang2015interface} and liquid-gating technique \cite{lei2015evolution, hanzawa2015field} raise new concerns over the superconductivity in FeSe-related materials. Considering alkali-metals and small molecules interacted FeSe compounds with similarly high $T_\textrm{c}$ ($\sim$ 40 K) as well \cite{Guo2010superconductivity, burrard2013enhancement, lu2014coexistence}, one central issue which naturally arises is whether the heavily electron-doped FeSe compounds including single-layer FeSe/SrTiO$_3$ represent novel superconductors with completely distinct pairing mechanism (e.g.\ phonon-mediated electron paring) from Fe-SCs, or whether they are merely some derivatives of heavily electron-doped Fe-SCs. In order to address this question, it is highly tempting to study systematically how the superconductivity is altered from a low-$T_\textrm{c}$ phase in undoped parent FeSe to high-$T_\textrm{c}$ phase in heavily electron-doped FeSe with increasing electron doping level $x$. However, all previous attempts at establishing such phase diagram were conducted either in nonsuperconducting strained multilayer FeSe/SrTiO$_3$ films \cite{miyata2015high, wen2015anomalous, tang2015interface}, or in liquid-gating tuned FeSe thin flakes suffering from significant inhomogeneity of electron-density distribution \cite{lei2015evolution}, which have severely hampered the clear identification of FeSe superconducting phase diagram.

Herein we report on such a phase diagram by exploring superconductivity in thickness-controlled FeSe ultra-thin films grown on graphitized SiC(0001) substrate \cite{song2011direct, song2011molecular, song2015imaging} with increasing surface K dosage using scanning tunneling microscopy/spectroscopy (STM/STS). This allows for a direct probing of superconducting order parameter at the nanoscale, thus avoiding macroscopic integral measurements involved in ARPES and transport techniques \cite{miyata2015high, wen2015anomalous, lei2015evolution}. Meanwhile, the nearly ``free-standing'' FeSe films on graphitized SiC(0001) \cite{song2011molecular} rule out possible interruption of the lattice mismatch-caused epitaxial strain and zebra-like stripes in multilayer FeSe/SrTiO$_3$ films \cite{tan2013interface, peng2014measurement, li2014molecular}. Our experiments were carried out on a Unisoku ultrahigh vacuum cryogenic STM system equipped with a molecular beam epitaxy (MBE) for \textit{in-situ} FeSe film growth. High-quality superconducting FeSe/SiC(0001) thin films with varying thickness were prepared following the well-established co-deposition method, as described in our previous reports \cite{song2011direct, song2011molecular}. K atoms were then evaporated from a well-outgassed SAES getter source onto FeSe films kept at $\sim$ 150 K. Prior to the STM/STS measurements at 4.3 K, polycrystalline PtIr tips were cleaned by electron-beam heating and then calibrated on MBE-grown Ag/Si(111) films. Tunneling spectra were acquired using a standard lock-in technique with a small bias modulation of 0.2 mV at 966 Hz, unless other specified.

\end{spacing}

\begin{figure}[h]
\includegraphics[width=\columnwidth]{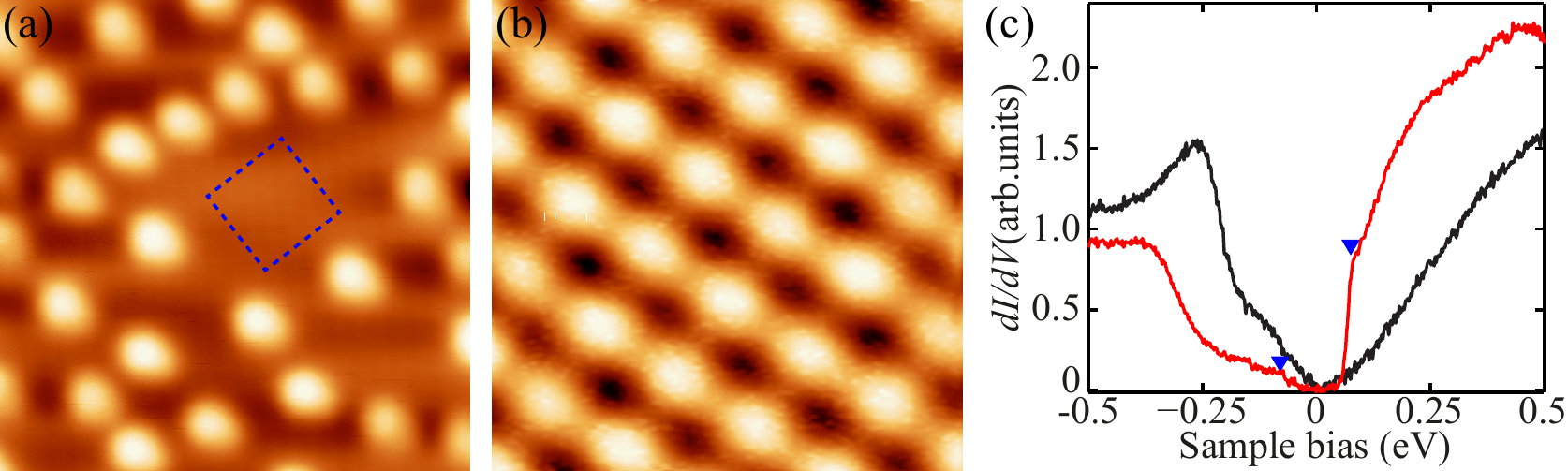}
\caption{(color online) (a) STM topographic image of K-doped FeSe/SiC(0001) films (10 nm $\times$ 10 nm, $V$ = 1.0 V, $I$ = 50 pA). (b) Atomically resolved topography measured in K-free region (2 nm $\times$ 2 nm, $V$ = 2 mV, $I$ = 100 pA), marked by the dashed square in (a). (c) Differential conductance $dI/dV$ spectra of FeSe films before (black curve) and after (red curve) a surface dose of 0.205 ML K. The two triangles denote the edges of the accidental gap between electron and hole pockets around $\Gamma$ point. Tunneling gap is set at $V$ = 0.5 V, $I$ = 100 pA. The lock-in bias modulation has a magnitude of 10 meV.
}
\end{figure}

Figure 1(a) depicts a constant-current topographic image of K doped FeSe/SiC(0001) films with a nominal K dosage of about 0.038 monolayer (ML). Here 1 ML is defined as the Se atomic number density at the topmost Se layer ($\sim 7 \times 10^{14}/\textrm{cm}^2$). Evidently, individual isolated K adatoms are randomly distributed at the surface. The absence of K dimer, multimer or cluster hints at strong repulsive interaction among the ionized K adatoms because of electron transfer from K to FeSe \cite{song2012charge}, consistent with previous reports \cite{miyata2015high, wen2015anomalous, tang2015interface}. Zoom-in STM images on any regions with sparse K adatoms (e.g.\ Fig.\ 1(b)) all reveal an untouched Se lattice, irrespective of how we post-anneal the samples at the elevated temperature ($<$ 400$^\textrm{o}$C). Plotted in Fig.\ 1(c) are the spatially averaged differential conductance $dI/dV$ spectra over a wide energy range (-0.5 eV $\sim$ 0.5 eV), acquired on FeSe before and after a surface dose of $\sim$ 0.205 ML K, respectively. The K adsorption does not lead to a simple rigid shift in the band structure of FeSe/SiC(0001) films, but instead suppresses strongly the spectral weight near the Fermi level ($E_F$). The resulted electronic structure resembles that of single-layer FeSe/SrTiO$_3$ films in a prominent manner, caused primarily by the accidental gapping between electron and hole pockets around $\Gamma$ point \cite{liu2012electronic, he2013phase, tan2013interface, lee2014interfacial, Huang2005revealing}. The gap size, defined as the energy separation between two gap edges (indicted by the blue triangles in Fig.\ 1(c)), is measured to be around 135 meV, quite close to the value of 140 meV in single-layer FeSe/SrTiO$_3$ films \cite{Huang2005revealing}. All these findings indicate that a systematic spectral survey of K doped FeSe/SiC films will shed some critical insights into superconductivity in FeSe-derived superconductors.

\begin{figure*}[t]
\includegraphics[width=2\columnwidth]{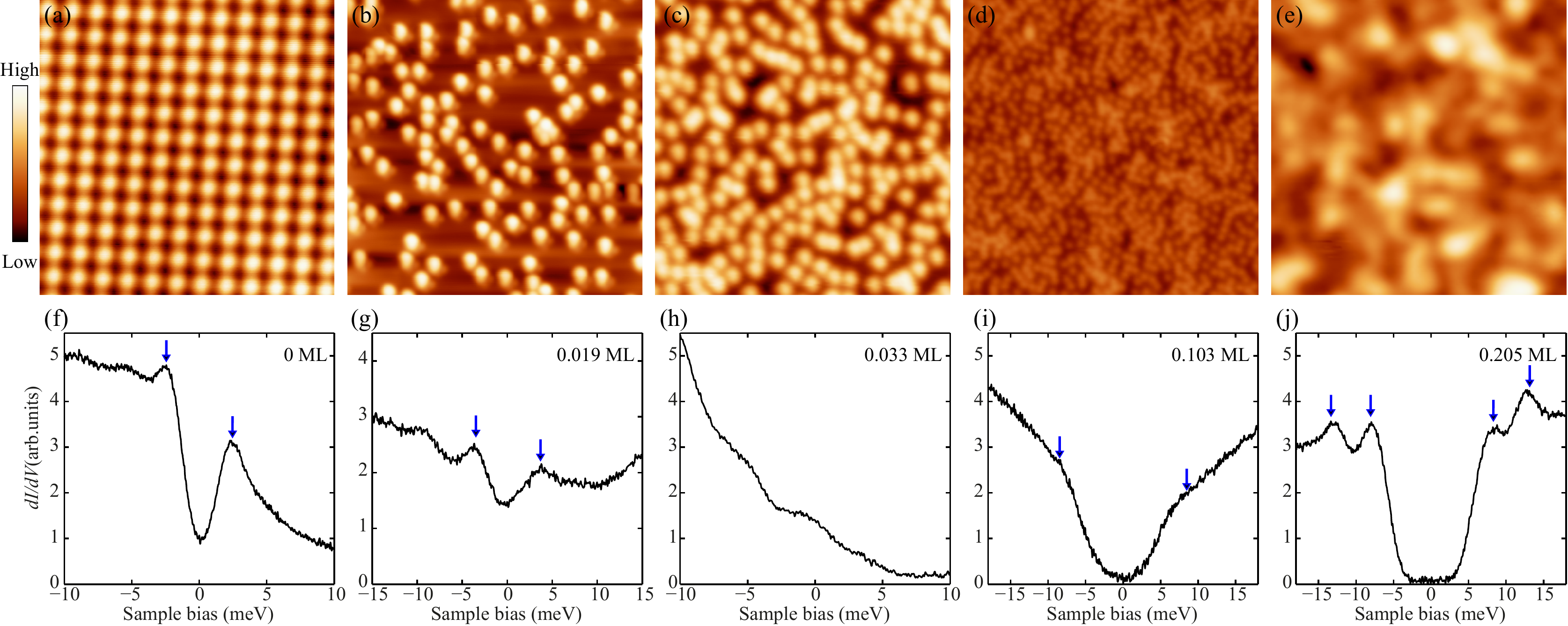}
\caption{(color online) (a-e) Topographies ((a) 5 nm $\times$ 5 nm; (b-e) 30 nm $\times$ 30 nm) and (f-j) $dI/dV$ spectra of multilayer FeSe/SiC(0001) films with varying doses of K, as indicated. Blue arrows denote the superconducting gap edges or coherence peaks. The absence of apparent $E_F$-symmetric gap in (h) shows full suppression of superconductivity in FeSe by an intermediate dose of K. Tunneling conditions: (a) $V$ = -10 mV, $I$ = 100 pA; (b) $V$ = 4.0 V, $I$ = 20 pA; (c) $V$ = 4.0 V, $I$ = 10 pA; (d) $V$ = 1.0 V, $I$ = 10 pA; (e) $V$ = 3.0 V, $I$ = 20 pA; (f) $V$ = 10 mV, $I$ = 100 pA; (g-j) $V$ = 20 mV, $I$ = 100 pA.
}
\end{figure*}

Enumerated in Fig.\ 2 are a series of topographies and much smaller-energy-ranged $dI/dV$ spectra in multilayer FeSe/SiC films as the K dosage is gradually increased. As anticipated, the parent FeSe film in Fig.\ 2(a) exhibits a single dominant superconducting gap $\Delta$ $\sim$ 2.2 meV [Fig.\ 2(f)], matching exactly with previous studies \cite{song2011direct, song2011molecular, song2015imaging}. As the dose is increased, individual isolated K adatoms tend to pile up together [Figs.\ 2(d) and 2(e)], analogous to K coated FeSe/SrTiO$_3$ films \cite{tang2015interface, tang2015superconductivity}. Quite interestingly, a considerable amount of U-shaped spectral weight depletion or loss, with two sets of $E_F$-symmetric peaks (black arrows) at the higher energy positions of $\sim$ 14 meV and 8.5 meV, respectively, is invariably revealed in heavily K-doped FeSe/SiC films [Fig.\ 2(j)]. This bears a striking resemblance to those observed in heavily electron-doped FeSe-derived superconductors \cite{qing2012interface, miyata2015high, wen2015anomalous, tang2015interface, lei2015evolution, hanzawa2015field, Guo2010superconductivity, burrard2013enhancement, lu2014coexistence}, signalling the occurrence of high-$T_\textrm{c}$ superconductivity in heavily K doped multilayer FeSe/SiC films.

\begin{figure}[b]
\includegraphics[width=0.97\columnwidth]{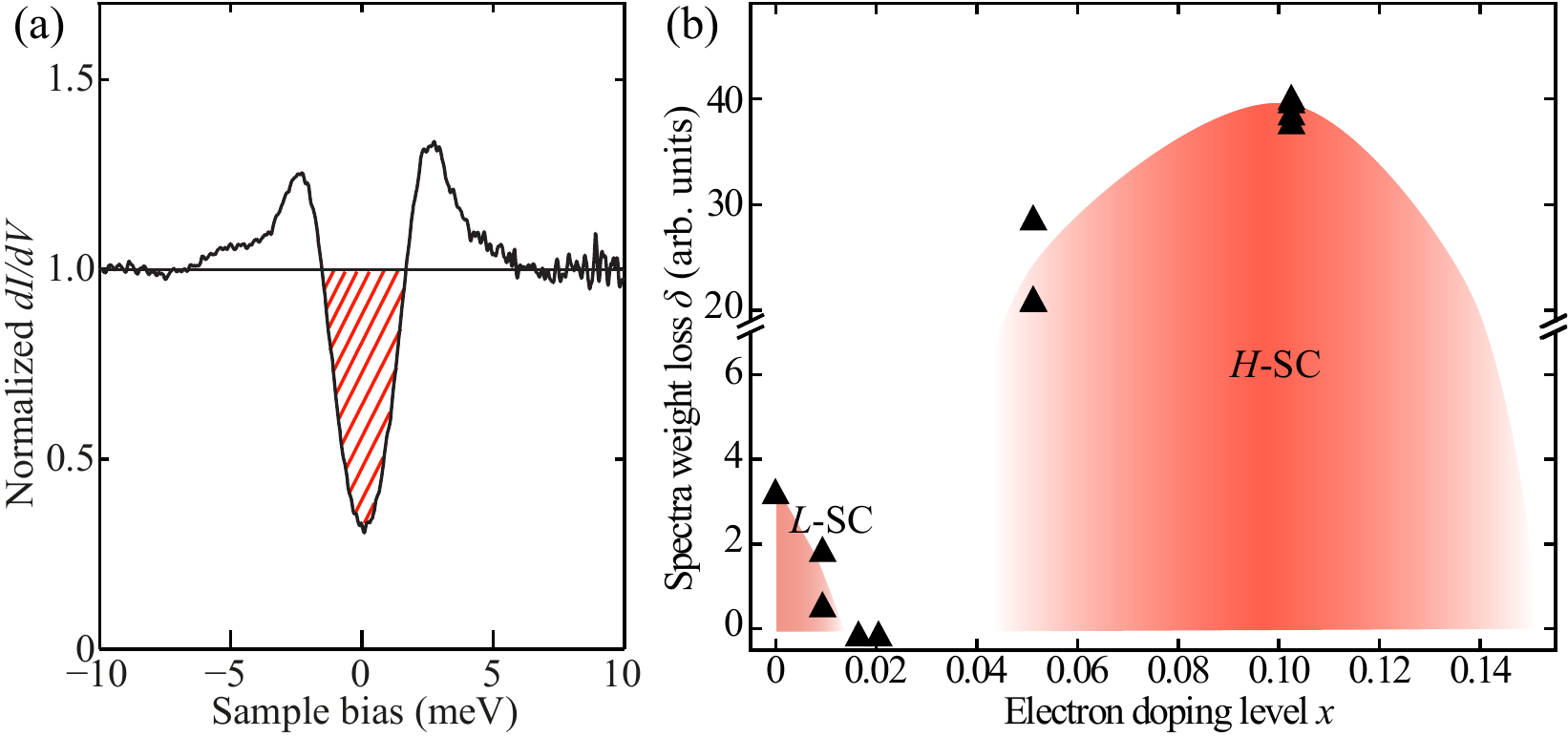}
\caption{(color online) (a) Normalized $dI/dV$ spectrum by dividing the raw $dI/dV$ spectrum in Fig.\ 2(f) by its background, which was extracted from a cubic fit to the conductance for $|V|>$ 6 mV. Red-shaded region characterizes the superconductivity-induced spectral weight loss $\delta$ near $E_F$. (b) Electron doping level $x$ dependence of $\delta$ (black triangles) or superconductivity, confirming two disconnected superconducting domes ($L$-SC and $H$-SC) in the electron-doped FeSe phase diagram.
}
\end{figure}

Furthermore, we found that with increasing K dosage, the stoichiometric parent FeSe with a low $T_\textrm{c}$ (or equivalently small $\Delta$) does not evolve monotonically into the high-$T_\textrm{c}$ phase, rather its superconductivity weakens firstly [Fig.\ 2(g)], vanishes entirely [Fig.\ 2(h)] and re-emerges abruptly with an enhanced gap magnitude $\Delta$ (thus high-$T_\textrm{c}$ superconductivity) in heavily K doped FeSe/SiC films [Figs.\ 2(i) and 2(j)]. To characterize this tendency more visibly, more $dI/dV$ spectra have been measured at various K doped FeSe and normalized, with one of them illustrated in Fig.\ 3(a). Assuming that a K adatom doses one electron into the low-lying FeSe films, we can summarize the superconductivity-induced spectral weight loss $\delta$ (red-shaded region in Fig.\ 3(a)) as a function of electron doping level $x$ (electrons per Fe, namely half of K dosage) in Fig.\ 3(b). Here a large $\delta$ means the larger spectral weight loss due to the superconducting gap opening, and thus characterizes reasonably the superconductivity or $T_\textrm{c}$. An unexpected phase diagram with two disconnected superconducting phases and a generally wide nonsuperconducting valley in between is visible, in clear contrast to the single-dome phase diagram reported recently \cite{wen2015anomalous}. This constitutes the major finding in this study.

A double-dome superconducting phase diagram has previously been identified in alkali-metals and ammoniated metal-intercalated FeSe superconductors modulated by external pressure \cite{sun2012re, izumi2015emergence}, whose mechanism so far escapes a reasonable explanation. Notably, however, the double-dome superconducting phase diagram established here differs markedly from the previous ones in terms of the external control parameter (electron doping $vs$ pressure) and whether or not the two superconducting domes are well separated (yes $vs$ not). Moreover, the two previous studies began from the already heavily electron-doped high-$T_\textrm{c}$ phase, contrasting the present study where we start from a undoped parent FeSe with a low $T_\textrm{c}$ of $<$ 10 K. Therefore, the observation of two well-disconnected superconducting domes here, tuned by electron doping level $x$, is intriguing and constitutes a novel experimental basis for unraveling the secret of Cooper paring in FeSe-derived superconductors.

\begin{figure}[b]
\includegraphics[width=\columnwidth]{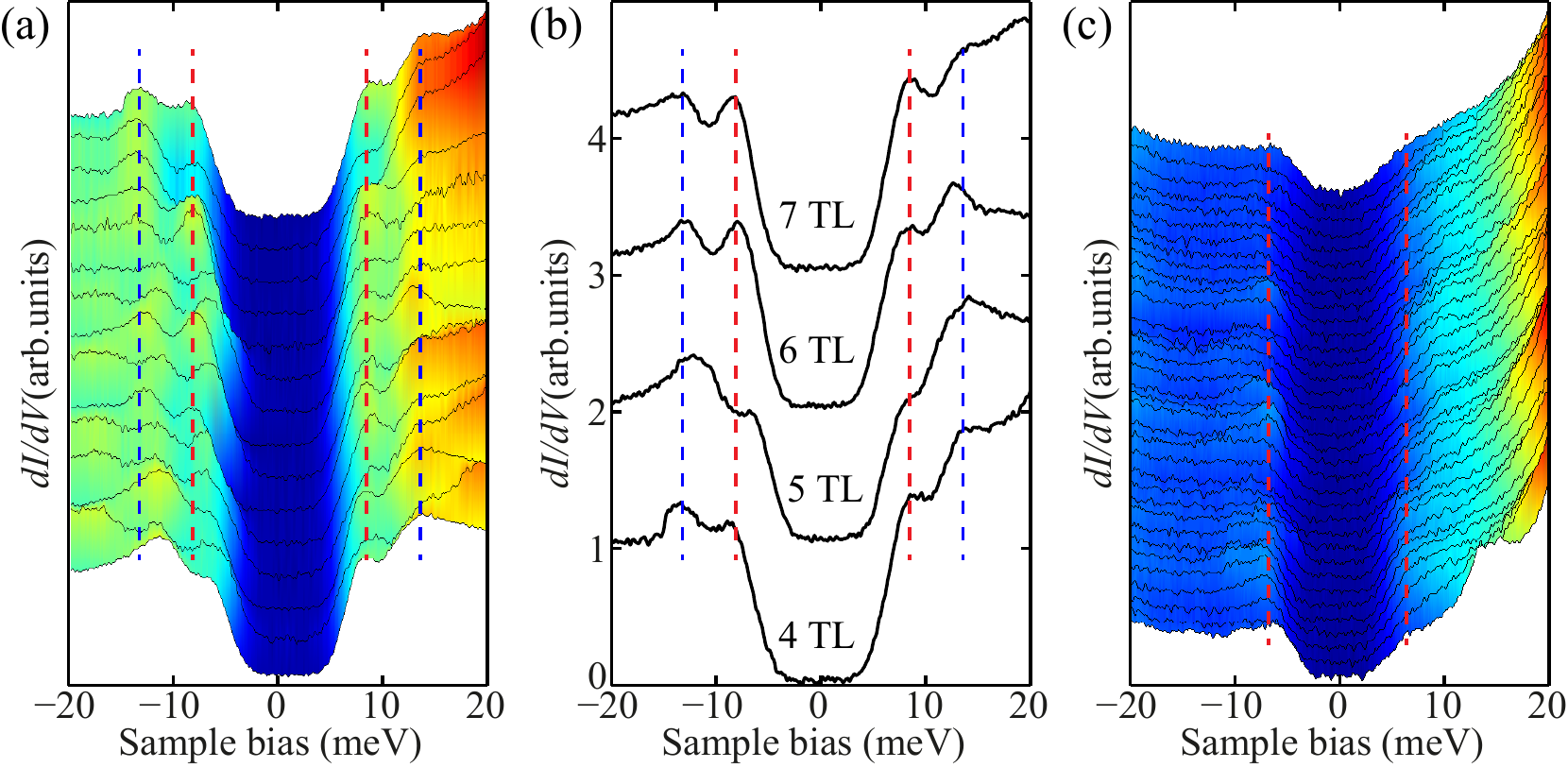}
\caption{(color online) (a) Spatial and (b) film thickness dependence of $dI/dV$ spectra acquired on heavily electron-doped multilayer FeSe films ($x \sim$ 0.103). Red and blue dashes show the approximate energy positions of two-energy-scale superconducting gaps, respectively. The film thickness of FeSe is indicated in a unit of triple layer (TL). (c) A series of $dI/dV$ spectra acquired along a 10-nm trajectory on heavily electron-doped single layer FeSe/SiC films ($x \sim$ 0.103). Red dashes indicate the energy positions of the superconducting gap. Setpoint: $V$ = 20 mV, $I$ = 100 pA.
}
\end{figure}

Now the identification of two disconnected superconducting domes raises the most important concern as to whether the Cooper coupling interaction is of the same mechanism in the two disconnected superconducting domes. To bring insight into this question as well as the pairing nature of heavily electron-doped FeSe high-$T_\textrm{c}$ superconductors, we have explored spatial and film thickness dependence of superconductivity in heavily electron-doped FeSe films, respectively. Figure 4(a) typifies a series of $dI/dV$ spectra at various sites of K doped FeSe/SiC films, which all exhibit the two-gap feature and prove spatially homogeneous. This implies that the electrons injected by surface K are mostly itinerant other than localized in the \textit{ab} plane. Additionally, the two-gap structure and gap magnitude $\Delta$ rely little on film thickness [Fig.\ 4(b)], unless that the film is thinned down to the two-dimensional limit, namely single-layer FeSe with a smaller single dominant superconducting gap of about 6.6 meV [Fig.\ 4(c)]. This exception might be caused by the enhanced thermal/quantum fluctuations in free-standing single-layer FeSe/SiC films, which weaken the superconductivity there \cite{tinkham2012introduction}. Nevertheless, our finding suggests an exclusive high-$T_\textrm{c}$ superconductivity at the topmost FeSe layer. This goes in line with non-intercalation behavior of K in FeSe/SiC films [Fig.\ 1(b)], leaving the layers beneath unchanged and thus avoiding interference of other unwanted phase, e.g.\ K$_x$Fe$_2$Se$_2$ in K doped multilayer FeSe/SrTiO$_3$ films \cite{miyata2015high, wen2015anomalous, tang2015interface, tang2015superconductivity}. The gap magnitude of $\sim$ 14 meV observed appears larger than most values reported in K doped multilayer FeSe/SrTiO$_3$ films \cite{miyata2015high, wen2015anomalous, tang2015interface}, but still smaller than $\Delta$ $\sim$ 20 meV in single-layer FeSe/SrTiO$_3$ \cite{qing2012interface}. Considering the smaller $\Delta$ $\sim$ 6.6 meV in K doped single-layer FeSe/SiC films [Fig.\ 4(c)], our experiment corroborates that the SrTiO$_3$ substrate plays more than electron doping to boost the high-$T_\textrm{c}$ superconductivity in, or only in single-layer FeSe/SrTiO$_3$ films, e.g.\ via interfacial phonon-enhanced pairing strength \cite{qing2012interface, lee2014interfacial, cui2015interface, lee2015makes}.

The rather robust superconductivity [Fig.\ 4(a)] against severe disorder at the heavily K doped FeSe surface (see Fig.\ 2(e)), in combination with the consistent revelations of plain-\textit{s} wave electron pairing in heavily electron-doped FeSe-derived superconductors \cite{fan2015plain, yan2015surface}, contradicts with most unconventional sign-changing pairing symmetry. Instead the findings comply with a conventional phonon-based Cooper pairing mechanism in this category of superconductors including single-layer FeSe/SrTiO$_3$ \cite{Balatsky2006impurity}. This is more evidenced by the vanishing spin fluctuation in Li$_x$(C$_2$H$_8$N$_2$)$_y$Fe$_{2-z}$Se$_2$ \cite{hrovat2015enhanced} and the recent tunneling measurement of FeSe Eg(Se) phonon mode ($\sim$ 11 meV) in K doped multilayer FeSe/SrTiO$_3$ \cite{tang2015interface}, although further research efforts are needed to fully pin down this issue. Provided that the spin fluctuation picture works in parent FeSe, two distinct paring mechanisms thus appear to operate in electron-doped FeSe, and the heavily electron-doped ones including single-layer FeSe/SrTiO$_3$ are a novel class of superconductors with completely different electron pairing mechanism from Fe-SCs, reflected more cogently by the contrasting properties between them. These, for examples, consists of the superconducting gap structure \cite{song2011direct, qing2012interface, fan2015plain, yan2015surface}, nematic \cite{song2011direct, Nakayama2014reconstruction, fan2015plain, yan2015surface, Huang2015nanoscale} and spin fluctuations \cite{Imai2009why, hrovat2015enhanced}. By contrast, even though the two disconnected superconducting phases are of the identical Cooper pairing mechanism, the absence of spin and nematic fluctuations in $H$-SC phase conversely supports that neither spin nor nematic fluctuations are the essential ingredient giving rise to high-$T_\textrm{c}$ superconductivity \cite{fan2015plain, yan2015surface, Huang2015nanoscale, hrovat2015enhanced, jie2015new}.

Finally we comment on why the multilayer FeSe films grown on SrTiO$_3$ substrate exhibit no superconductivity before K doping, a long-standing mystery confusing the FeSe superconductivity community \cite{qing2012interface}. As seen from Fig.\ 3(b), the superconductivity is hypersensitive to electron doping in the lower-$T_\textrm{c}$ phase ($L$-SC), and an injection of only $\sim$ 0.015 electron/Fe into FeSe can completely kill its superconductivity. As thus, a small amount of but insufficient electron doping from SrTiO$_3$ substrate to multilayer FeSe films will push them to the nonsupercondcuting valley between the $L$-SC and $H$-SC phases [Fig.\ 3(b)]. In support of this standpoint, we have conducted a comparative study, and found that the multilayer FeSe/SrTiO$_3$ films need less K dose to gain high-$T_\textrm{c}$ superconductivity than FeSe/SiC films \cite{tang2015interface}. This indicates that the nonsuperconducting multilayer FeSe/SrTiO$_3$ films have indeed been electron doped as compared to parent FeSe.

Our detailed real-space STM/STS scrutiny of K doped FeSe/SiC films has demonstrated a high-$T_\textrm{c}$ superconductivity at the topmost FeSe layer, and how the superconductivity evolves from the low-$T_\textrm{c}$ phase in parent FeSe to high-$T_\textrm{c}$ phase in heavily electron-doped FeSe superconductors. The emergence of two disconnected superconducting domes is beyond expectation and will certainly stir up a number of further experimental and theoretical studies. Our work places a severe constraint on the theoretical model for understanding superconductivity in FeSe-related superconductors.

\begin{acknowledgments}
This work was financially supported by National Science Foundation and Ministry of Science and Technology of China. C. L. S acknowledges support from Tsinghua University Initiative Scientific Research Program. %The STM images were processed using WSxM software \cite{horcas2007wsxm}.
\end{acknowledgments}

%\bibliography{KFeSe}
%

\end{document}